\documentclass[a4paper,12pt]{article}
\usepackage[pctex32]{graphics}
\usepackage{amssymb,amsmath}
\usepackage{amsmath,amssymb}
\usepackage{latexsym}
\usepackage{epsfig}
\usepackage[english]{babel}

\newcommand{\be}{\begin{equation}}
\newcommand{\ee}{\end{equation}}
\newcommand{\ba}{\begin{eqnarray}}
\newcommand{\ea}{\end{eqnarray}}

\begin{document}

\begin{titlepage}

\vspace{5mm}

\begin{center}

{\Large \bf The origin of  regular  Newtonian  \\potential   in  infinite derivative   gravity}

\vskip .6cm

{Yun Soo Myung$^a$\footnote{e-mail address: ysmyung@inje.ac.kr} and Young-Jai Park$^b$\footnote{e-mail address:yjpark@sogang.ac.kr}}\\[8mm]

{$^a$Institute of Basic Sciences and Department  of Computer Simulation, \\Inje University Gimhae 50834, Korea\\[0pt]}
{$^b$Department of Physics, Sogang University, Seoul 04107, Korea\\[0pt]}

\vspace{5mm}

\end{center}

\begin{center}
\underline{Abstract}
\end{center}
It turns out that the infinite derivative gravity (IDG) is ghost-free and renormalizable when one chooses the exponential of an  entire function.
For this IDG case,  the corresponding Newtonian potential generated from the delta function   is non-singular at the origin.
However, we will explicitly show  that the source generating  this non-singular potential  is  given not by the delta-function due to the point-like source of mass, but by the Gaussian mass distribution.
This explains clearly  why  the IDG with the exponential of an  entire function  yields  the finite potential  at the origin.
 \vskip .6cm

\vskip 0.8cm

\vspace{15pt} \baselineskip=18pt

\thispagestyle{empty}
\end{titlepage}

\newpage
%%%%%%%%%%%%%%%%%%%%%%%%%%%%%%%%%%%%%%%%%%%%%%%%%%%%%%%%%%%%%%%%%%
\section{Introduction}
There was a conjecture that renormalizable higher-derivative gravity has  a finite Newtonian potential at the  origin~\cite{Accioly:2013hwa,Modesto:2014eta,Giacchini:2016xns,Accioly:2017xmm}.
This relation was first mentioned  in Stelle's seminal work~\cite{Stelle:1976gc} which showed that the fourth-derivative gravity  is renormalizable and  has a finite potential at origin.
Unfortunately, this gravity suffers from ghost problem  because it has a massive pole with negative residue which could be interpreted  either as a state of negative norm or a state of negative energy.

It turned out that  the infinite derivative gravity (IDG) can resolve the problem of massive ghost  as well as it may avoid the singularity of the Newtonian potential at the origin, when one chooses the exponential of an  entire function. We would like to note that this model is also named super-renormalizable quantum gravity~\cite{Modesto:2011kw}.
However, one does not understand fully  how this nonlocal  gravity could provide a regular potential.
It was  argued that the cancellation of the singularity at the origin  is an effect of an infinite amount of hidden ghost-like complex poles~\cite{Modesto:2014eta}.

On the other hand, Tseytlin~\cite{Tseytlin:1995uq} showed already  in string theory that the corrected Poisson equation [$e^{-c\alpha' \Delta}\Delta K=-\mu\delta^{(N)}(x)$] with $\Delta=\delta_{ij}\partial_i\partial_j$ regularizes effectively the delta-function source at the scale of $\sqrt{\alpha'}$.
In other words, replacing  the source $\delta^{(N)}(x)$ by  $\delta^{(N)}_{\alpha'}(x)=e^{c\alpha'\Delta}\delta^{(N)}$, one can interpret ``$ \Delta K=-\mu\delta^{(N)}_{\alpha'}(x)$" as
a Poisson equation with  smearing of the delta-function source at the string scale $\sqrt{\alpha'}$.
Furthermore, the source of $\delta^{(N)}_{\alpha'}(x)$ removes  Newtonian singularity~\cite{Shapiro:2015uxa}, similar to the simpler fourth-derivative gravity~\cite{Stelle:1976gc}.
Especially, we would like to point out    that the Gaussian mass distribution gives rise to  a regular Newtonian potential in the Einstein gravity~\cite{Casadio:2017cdv}

In this work,  we will  show  that the source generating  the  regular Newtonian  potential  is  originated  not from  the delta-function due to the point-like mass source but from the Gaussian mass distribution.
This indicates clearly  why  the IDG with the exponential of an  entire function  yields  the finite potential  at the origin.
Transforming  the delta-function source (Einstein theory) to the Gaussian distribution (IDG) makes the Newtonian potential regular.

\section{Propagator for IDG  }
In this section, we wish to mention briefly   a process for  obtaining a non-singular Newtonian potential  from the propagator of the IDG.
Starting from the general version  of higher-derivative gravitational action up to the second order in curvature without restricting the number of derivative
\begin{equation} \label{idg-0}
S=\frac{1}{4\kappa^2}\int d^4x\sqrt{-g} \Big[-2R+RF_1(\square_\sigma)R+R_{\mu\nu}F_2(\square_\sigma)R^{\mu\nu}+R_{\mu\nu\rho\eta}F_3(\square_\sigma)R^{\mu\nu\rho\eta}\Big],
\end{equation}
where $\sigma$ is the length scale at which the nonlocal modifications become important. Here $F_i$'s are infinite-derivative functions of $\square_\sigma$ [$F_i(\square_\sigma)=\sum_{n=0}c_{i_n}\square_\sigma^n$].
One needs a specific constraint of $2F_1+F_2+2F_3=0$~\cite{Biswas:2011ar,Biswas:2013kla} when expanding around a Minkowski spacetime ($g_{\mu\nu}=\eta_{\mu\nu}+ h_{\mu\nu})$ with $\eta_{\mu\nu}={\rm diag}(+---)$ so that the bilinear action can provide a ghost-free (massive) tensor propagator.
The simplest condition could be achieved if one chooses $F_i$'s as
\begin{equation}
F_1(\square_\sigma)=\frac{a(\square_\sigma)-1}{\square_\sigma},~~F_2(\square_\sigma)=-2F_1(\square_\sigma),~~F_3(\square_\sigma)=0,
\end{equation}
which corresponds to the relation of $a(\square_\sigma)=c(\square_\sigma)$ with
\begin{equation}
a(\square_\sigma)=1- \frac{\square_\sigma F_2(\square_\sigma)}{2},~c(\square_\sigma)=1+\square_\sigma \Big(F_2(\square_\sigma)+3F_1(\square_\sigma)\Big).
\end{equation}
This implies, for non-linear case, that a simplest model of the IDG is given by~\cite{Modesto:2014eta}
\begin{equation} \label{idg-1}
S_{\rm IDG}=-\frac{1}{2\kappa^2}\int d^4x\sqrt{-g} \Big[R+G_{\mu\nu}\frac{a(\square_\sigma)-1}{\square} R^{\mu\nu}\Big]
\end{equation}
with $\kappa^2=8\pi G$.
Bilinearizing the Lagrangian of  Eq.(\ref{idg-1})  together with  imposing the de Donder gauge,
one obtains ${\cal L}^{\rm bil}_{\rm IDG}=h^{\mu\nu}{\cal O}_{\mu\nu,\alpha\beta}h^{\alpha\beta}/2$. The inverse operator ($1/{\cal O}$) corresponds to the propagator.
The propagator of the action (\ref{idg-1}) takes the form with $k^2=k^2_0-{\bf k}^2({\mathbf k}^2={\mathbf k}\cdot {\mathbf k}=|{\mathbf k}|^2) $
\begin{equation}\label{prop-to}
{\cal D}^{\rm IDG}_{\mu\nu,\rho\sigma}(k)=\frac{1}{ a(-k^2_\sigma)}\frac{1}{k^2}\Big[P^{(2)}_{\mu\nu,\rho\sigma}-\frac{P^{(0-s)}_{\mu\nu,\rho\sigma}}{2}\Big],
\end{equation}
where $P^{(2)}$ and $P^{(0-s)}$ are the Barnes-Rivers spin projection operators.
If one chooses $a(\square_\sigma)=e^{\sigma^2\square/4}[a(-k^2_\sigma)=e^{-\sigma^2k^2/4}]$ which is the exponential of an entire function, there is no room to introduce ghost poles
at the perturbative level for the IDG.
This is because $a(\square_\sigma)$ has no zeros or no poles. In this case,
the only dynamical pole resides at $k^2=0$ which corresponds to  a massless  pole of  the spin-2 propagator.
Given this entire function, furthermore, the propagator becomes exponentially suppressed in the UV and the vertex factors are exponentially enhanced
so that this theory can be renormalizable.
We note that the other choice of  $a(\square_\sigma)=e^{-\sigma^2\square/4}[a(-k^2_\sigma)=e^{\sigma^2k^2/4}]$ with  $\eta_{\mu\nu}={\rm diag}(+---)$
does not leads to the Gaussian distribution in momentum space.

Let us pay attention to  the relation between static propagator of ${\cal D}^{\rm IDG}_{\mu\nu,\rho\sigma}(k)|_{k_0=0}$  and  classical potential generated by a point-like  mass source of
$T_{\mu\nu}=\rho\eta_{\mu 0}\eta_{\nu 0}$ with $\rho=M\delta^3({\bf r})$,
\begin{equation}
V(r)=\frac{\kappa^2 M}{(2\pi)^3}\int d^3\mathbf{k} e^{i\mathbf{ k}\cdot\mathbf{ r}} {\cal D}_{00,00}(\mathbf{k}),~~~~ {\cal D}_{00,00}^{\rm IDG}(\mathbf{k})=-\frac{e^{-\sigma^2 |{\mathbf k}|^2/4}}{2|{\mathbf k}|^2}.
\end{equation}
Then, we obtain the Newtonian  potential
\begin{equation} \label{newtp-1}
V^{\rm IDG}(r)=-\frac{GM}{r}{\rm Erf}\Big(\frac{ r}{\sigma}\Big).
\end{equation}
Since the error function  ${\rm Erf}(x)$ takes a series form around $x=0$ as
\begin{equation}
{\rm Erf}(x)\simeq \frac{2x}{\sqrt{\pi}} -\frac{x^3}{\sqrt{\pi}} +\frac{11x^5}{20\sqrt{\pi}} -\frac{241x^7}{840\sqrt{\pi}} +\cdots,
\end{equation}
 the regular potential is obtained from  Eq.(\ref{newtp-1}) as 
\begin{equation}\label{newtonp}
  V^{\rm IDG}(r) \simeq \frac{2GM}{\sqrt{\pi}\sigma}[-1+\frac{ r^2}{2\sigma^2}-\cdots].
\end{equation}
 As a result,  we find   that around $r=0$, the singularity disappears due to the non-locality
and the effect depends on a specific  form of $a(\square_\sigma)$.
Hence, the action (\ref{idg-1}) provides a ghost-free and singularity-free  gravity which is also renormalizable.

\section{Origin of regular potential}
At this stage, let us  remind that the Newtonian potential $V^{\rm IDG}(r)$ was  obtained from the dressed propagator (\ref{prop-to})  with the point-like source $\rho=M\delta^3({\bf r})$.
It was argued that the cancellation of singularity may be seen as an effect of an infinite amount of hidden ghost-like complex poles~\cite{Modesto:2014eta}.
However, we point out that  the dressed propagator has only a simple pole at $k^2=0$ because $a(-k^2_\sigma)$ is the exponential of an entire function.
Thus,  this could not explain  an origin of the regular potential (\ref{newtp-1}). This is so because one derived the potential from the propagator by assuming the point-like source.

Since the regular potential implies the renormalizable gravity, we must  explore the origin of the regular potential by investigating the mass source.
For this purpose, we derive the linearized  equation  from (\ref{idg-1}) with the external  source $T_{\mu\nu}$~\cite{Buoninfante:2016iuf}
\begin{equation} \label{lineq1}
a(\square_\sigma)\Big[\square h_{\mu\nu}-(\partial_\mu \partial_\alpha h^\alpha_\nu+\partial_\alpha \partial_\nu h^\alpha_\mu)+(\eta_{\mu\nu}\partial_\alpha \partial_\beta h^{\alpha\beta}+\partial_\mu\partial_\nu h)
-\eta_{\mu\nu}\square h\Big]=-2\kappa^2 T_{\mu\nu}.
\end{equation}
Considering the Newtonian approximation of $\partial_0 h_{\mu\nu}=0$,  the trace and  the  00-component of (\ref{lineq1}) take the forms
\begin{eqnarray}
&& 2a(\square_\sigma)\Big[-\square h+\partial_\alpha \partial_\beta h^{\alpha\beta}\Big]=-2\kappa^2 \rho, \\
&&a(\square_\sigma)\Big[\square h_{00}+\partial_\alpha \partial_\beta h^{\alpha\beta}-\square h\Big]=-2\kappa^2 \rho.
\end{eqnarray}
Choosing the perturbation as the Newtonian and Bardeen potentials as
\begin{equation}
h_{\mu\nu}={\rm diag}\{2\Phi,2\Psi,2\Psi,2\Psi\},
\end{equation}
the trace and the 00-component equations are given by
\begin{eqnarray} \label{zz-1}
&& 2a(\Delta_\sigma)\Big[2\Delta(\Phi-3\Psi)+2\Delta\Psi\Big]=-2\kappa^2 \rho, \\
 \label{zz-2}&&4a(\Delta_\sigma)\Delta\Psi=2\kappa^2 \rho.
\end{eqnarray}
By comparing (\ref{zz-1}) with (\ref{zz-2}), we find that two potentials $\Phi$ and $\Psi$ satisfy the same  equation
\begin{equation} \label{Npot-eq}
2a(\Delta_\sigma)\Delta\{\Phi,\Psi\}=\kappa^2 M \delta^3({\bf r}).
\end{equation}
We wish to find the solution to (\ref{Npot-eq}) by going  into the momentum space and then, going back to the coordinate space.
Fourier transforming  (\ref{Npot-eq}) leads to
\begin{equation} \label{fpeq-1}
-2a({\bf k}^2_\sigma){\bf k}^2\Phi({\bf k})=\kappa^2 M,~~a({\bf k}^2_\sigma)=e^{\frac{\sigma^2|{\bf k}|^2}{4}}
\end{equation}
whose potential  is determined  by inverse-Fourier transforming as
\begin{equation}
\Phi(r)=-\frac{\kappa^2 M}{2}\int \frac{ d^3{\bf k}}{(2\pi)^3} \frac{e^{i {\bf k}\cdot {\bf r}}}{{\bf k}^2 a({\bf k}^2_\sigma)}=
-\frac{\kappa^2 M}{(2\pi)^2}\frac{1}{r} \int^{\infty}_{0}d|{\bf k}|\frac{e^{-\sigma^2|{\bf k}|^2/4} \sin[|{\bf k}|r]}{|{\bf k}|}.
\end{equation}
Noting
\begin{equation} \label{newtp-2}
  \int^{\infty}_{0}d|{\bf k}|\frac{e^{-\sigma^2|{\bf k}|^2/4} \sin[|{\bf k}|r]}{|{\bf k}|}=\frac{\pi}{2} {\rm  Erf} \Big(\frac{r}{\sigma}\Big),
 \end{equation}
the Newtonian potential is given by
\begin{equation} \label{newtp-2}
\Phi_{\rm IDG}(r)=-\frac{GM}{r}{\rm Erf}\Big(\frac{ r}{\sigma}\Big),
\end{equation}
which is exactly the same form as in Eq.(\ref{newtp-1}).

Inspired by Tseytlin's work~\cite{Tseytlin:1995uq}, we transform (\ref{Npot-eq}) into the poisson-like equation
\begin{equation} \label{Npot-eq1}
\Delta\{\Phi,\Psi\} \equiv 4\pi G \rho(r)
\end{equation}
whose Fourier-transform is given  by~\cite{Casadio:2017cdv}
\begin{equation}
\Phi(|{\bf k}|)=-4\pi G \frac{\tilde{\rho}(|{\bf k}|)}{{\bf k}^2}.
\end{equation}
Here
\begin{equation} \label{rho-k}
\tilde{\rho}(|{\bf k}|)=4\pi \int^\infty_0 r^2 dr \frac{\sin[|{\bf k}|r]}{|{\bf k}|r} \rho(r).
\end{equation}

For example, given  a point-like source of mass $M$
\begin{equation} \label{rho-r}
\rho_{\rm p}(r)=M\delta^3({\bf r})=\frac{M}{4\pi r^2} \delta(r),
\end{equation}
 plugging (\ref{rho-r}) into (\ref{rho-k}) leads to the density in momentum space
\begin{equation} \label{rho-kk}
\tilde{\rho}_{\rm p}(|{\bf k}|)=M \int^\infty_0  dr \frac{\sin[|{\bf k}|r]}{|{\bf k}|r} \delta(r)=M.
\end{equation}
Performing inverse-Fourier transformation, one obtains  the usual Newtonian potential as
\begin{equation}
\Phi(r)_{\rm p}=-\frac{2G}{\pi} \int^\infty_0 d|{\bf k}|  \frac{\sin[|{\bf k}|r]}{|{\bf k}|r} \tilde{\rho}_{\rm p}(|{\bf k}|)=-\frac{2GM}{\pi r} \int^\infty_0 dz  \frac{\sin[z]}{z}=-\frac{GM}{r}.
\end{equation}
On the other hand, let us look for  the IDG-mass distribution whose momentum distribution is initially  given by (\ref{fpeq-1}) as
\begin{equation}
\tilde{\rho}_{\rm IDG}(|{\bf k}|)=Me^{-\sigma^2|{\bf k}|^2/4}.
\end{equation}
It is given by the inverse-Fourier transformation~\cite{Casadio:2017cdv}
\begin{equation}
\rho_{\rm IDG}(r)=\int^\infty_0\frac{|{\bf k}|^2d|{\bf k}|}{2\pi^2}  \frac{\sin[|{\bf k}|r]}{|{\bf k}|r} \tilde{\rho}_{\rm IDG}(|{\bf k}|)=\frac{M}{\pi^{3/2}}\frac{e^{-\frac{r^2}{\sigma^2}}}{\sigma^3},
\end{equation}
which is obviously the Gaussian mass distribution (regular matter density) with the width $\sigma$ when  comparing to the point-like source (\ref{rho-r}).
We check that the total mass is given by
\begin{equation}
M=4\pi \int^\infty_0 r^2 dr \rho_{\rm IDG}(r).
\end{equation}
Fig. 1 depicts the Gaussian mass density $\rho_{\rm IDG}(r)$, the IDG regular potential $\Phi_{\rm IDG}(r)$, and the singular Newtonian  potential  $\Phi_{\rm p}(r)$ as functions of $r$.
We observe that the density $\rho_{\rm IDG}(r)$ is essentially zero for $r\geq 3\sigma$, which allows us to identify   $\sigma$ as the length scale of the nonlocality.
Here, one finds that $\Phi_{\rm p}(r) \simeq \Phi_{\rm IDG}$ for $r\geq 3\sigma$.
\begin{figure*}[t!]
\centering
\includegraphics[width=.8\linewidth,origin=tl]{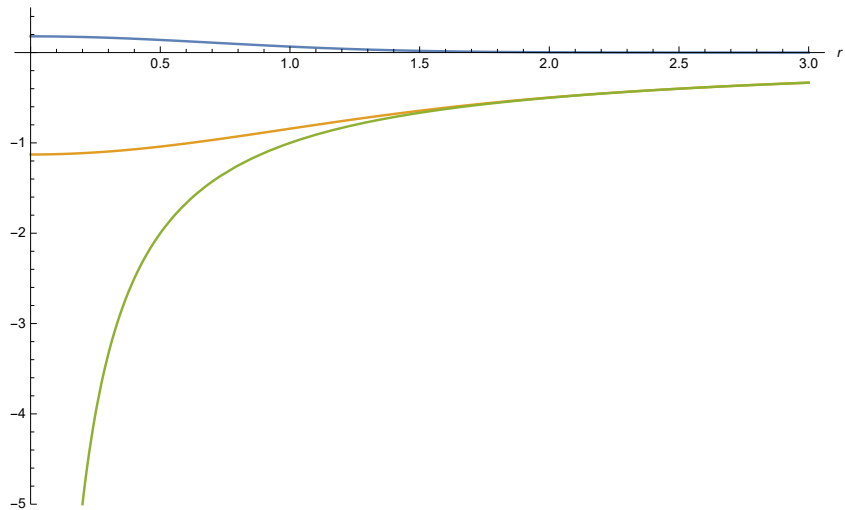}
\caption{For Gaussian mass distribution with $\sigma=1$ and $M=1$ (top curve), its regular Newtonian potential appears as middle curve. The bottom curve denotes singular Newtonian potential  for point-like source of mass $M=1$ with $G=1$. }
\end{figure*}

\section{Discussions}

Even though the nonlocal gravity of (\ref{idg-1}) with $a(\square_\sigma)=e^{\sigma^2\square/4}$ provides the regular Newtonian potential (\ref{newtonp}),
one did not know explicitly  what makes the potential finite at the origin.  On the other hand, the regular potential becomes an  important issue because it implies the renormalizable quantum gravity.

In this work, we have shown that Einstein gravity (Poisson equation)  provides the singular potential as shown in the process of $\rho_{\rm p}({\bf r})=M\delta^3({\bf r})\to \tilde{\rho}_{\rm p}({\bf k})=M \to \Phi_{\rm p}(r)=-GM/r$,
whereas the infinite derivative gravity (IDG) with $a(\square_\sigma)=e^{\sigma^2\square/4}$ (Poisson-like equation) gives us the regular potential as shown in the process of $\tilde{\rho}_{\rm IDG}({\bf k})=Me^{-\sigma^2|{\bf k}|^2/4}\to \rho_{\rm IDG}({\bf r})=(M/\pi^{3/2}\sigma^3)e^{-r^2/\sigma^2}\to \Phi_{\rm IDG}(r)=-GM{\rm Erf}(r/\sigma)/r$.

 Importantly, 
 we found the Gaussian mass distribution $\rho_{\rm IDG}({\bf r})$ from the Gaussian distribution $\tilde{\rho}_{\rm IDG}({\bf k})$ in momentum space. 
It implies that the origin of the regular potential in the IDG theory is the smearing of the point-like mass source for the nonlocal scale of $r<3\sigma$, corresponding to the Gaussian mass distribution.
For $r\ge 3\sigma$, one finds that two potentials are nearly the same like as $\Phi_{\rm p}(r) \simeq \Phi_{\rm IDG}$.
This explains clearly  why  the IDG with the exponential of an  entire function  yields  the finite potential  at the origin.

\section*{Acknowledgement}
This work was supported by the National Research Foundation of Korea (NRF) grant funded by the Korea government (MOE)
(No. NRF-2017R1A2B4002057).

\end{document}